\documentclass[preprint]{aastex}
\usepackage{color,natbib,lscape}

\newcommand{\Rsun}{\mbox{$R_{\sun}$}}

\newcommand{\Rjup}{\mbox{$R_{\rm Jup}$}}

\newcommand{\etal}{et al.}
\newcommand{\eg}{e.g.}

\newcommand{\kms}{\hbox{km~s$^{-1}$}}

\newcommand{\htwoo}{{\hbox{H$_2$O}}}   
\newcommand{\meth}{{\hbox{CH$_4$}}}   
\newcommand{\ammonia}{{\hbox{NH$_3$}}}   

\newcommand{\Ks}{\mbox{$K_S$}}
\newcommand{\degs}{\mbox{$^{\circ}$}}

\newcommand{\Teff}{\mbox{$T_{\rm eff}$}}
\newcommand{\logg}{\mbox{$\log(g)$}}

\newcommand{\PS}{\protect\hbox{Pan-STARRS1}}
\newcommand{\ps}{\protect\hbox{Pan-STARRS1}}
\newcommand\TPS{$3\pi$~Survey}
\newcommand{\gps}{\ensuremath{g_{\rm P1}}}
\newcommand{\rps}{\ensuremath{r_{\rm P1}}}
\newcommand{\ips}{\ensuremath{i_{\rm P1}}}
\newcommand{\zps}{\ensuremath{z_{\rm P1}}}
\newcommand{\yps}{\ensuremath{y_{\rm P1}}}

\newcommand{\grizy}{\gps, \rps, \ips, \zps, \yps}

\newcommand{\WISE}{{\sl WISE}}

\newcommand{\vtan}{\mbox{$V_{tan}$}}

\def\fdeg{\hbox{$.\!\!^\circ$}}            

\newcommand{\psobject}{\hbox{PSO~J043.5+02}}
\newcommand{\psobjectfull}{\hbox{PSO~J043.5395+02.3995}}
 
\newcommand{\wiseobjectfull}{\hbox{WISEP~J025409.45+022359.1}}

\slugcomment{ApJL, submitted 07/21/2011, revised 08/30/2011}

\shorttitle{PS1+2MASS+WISE High Proper Motion Search}
\shortauthors{Liu et al.}

\begin{document}

\title{A Search for High Proper Motion T~Dwarfs with\\
  Pan-STARRS1 + 2MASS + WISE}


\author{Michael C. Liu,\altaffilmark{1,2}
Niall R. Deacon,\altaffilmark{1}
Eugene A. Magnier,\altaffilmark{1}
Trent J. Dupuy,\altaffilmark{3,4}
Kimberly M. Aller,\altaffilmark{1}
Brendan P. Bowler,\altaffilmark{1}
Joshua Redstone,\altaffilmark{5}
Bertrand Goldman,\altaffilmark{6}
W. S. Burgett,\altaffilmark{1}
K. C. Chambers,\altaffilmark{1} 
K. W. Hodapp,\altaffilmark{1}
N. Kaiser,\altaffilmark{1}
R.-P. Kudritzki,\altaffilmark{1}
J. S. Morgan,\altaffilmark{1}
P. A. Price,\altaffilmark{7}
J. L. Tonry,\altaffilmark{1}
R. J. Wainscoat\altaffilmark{1}
}

\altaffiltext{1}{Institute for Astronomy, University of Hawaii, 2680
  Woodlawn Drive, Honolulu HI 96822 USA}
\altaffiltext{2}{Visiting Astronomer at the Infrared Telescope Facility,
  which is operated by the University of Hawaii under Cooperative
  Agreement no. NNX-08AE38A with the National Aeronautics and Space
  Administration, Science Mission Directorate, Planetary Astronomy
  Program.}
\altaffiltext{3}{Harvard-Smithsonian Center for Astrophysics, 60 Garden
  Street, Cambridge, MA 02138 USA}
\altaffiltext{4}{Hubble Fellow}
\altaffiltext{5}{Facebook, 1601 S. California Avenue, Palo Alto, CA
  94304, USA}
\altaffiltext{6}{Max Planck Institute for Astronomy, Koenigstuhl 17,
  D-69117 Heidelberg, Germany}
\altaffiltext{7}{Department of Astrophysical Sciences, Princeton University, Princeton, NJ 08544, USA}

\begin{abstract} 
  \noindent We have searched $\approx$8200 sq.~degs for high proper
  motion ($\approx$0.5--2.7\arcsec/year) T~dwarfs by combining
  first-epoch data from the Pan-STARRS1 (PS1) \TPS, the 2MASS All-Sky
  Point Source Catalog, and the \WISE\ Preliminary Data Release.
  We identified two high proper motion objects with the very red
  $(W1-W2)$ colors characteristic of T~dwarfs, one being the known
  T7.5~dwarf GJ~570D. 
  Near-IR spectroscopy of the other object
  (\psobjectfull~$\equiv$~\wiseobjectfull) reveals a spectral type of
  T8, leading to a photometric distance of $7.2\pm0.7$~pc. The
  2.56\arcsec/yr proper motion of \psobject\ is the second highest among
  field T~dwarfs, corresponding to an tangential velocity of
  $87\pm8$~\kms. According to the Besan\c{c}on galaxy model, this
  velocity indicates its galactic membership is probably in the thin
  disk, with the thick disk an unlikely possibility. Such membership is
  in accord with the near-IR spectrum, which points to a surface gravity
  (age) and metallicity typical of the field population.
  We combine 2MASS, SDSS, \WISE, and PS1 astrometry to derive a
  preliminary parallax of $171\pm45$~mas ($5.8^{+2.0}_{-1.2}$~pc), the
  first such measurement using PS1 data.
  The proximity and brightness of \psobject\ will facilitate future
  characterization of its atmosphere, variability, multiplicity,
  distance, and kinematics. 
  The modest number of candidates from our search suggests that the
  immediate ($\sim$10~pc) solar neighborhood does not contain a large
  reservoir of undiscovered T~dwarfs earlier than about T8.
\end{abstract}


\section{Introduction}

The \ps\ (PS1) survey \citep{PS1_system} represents the first of a new
generation of multi-epoch digital sky surveys. In an effort to identify
nearby T~dwarfs and rare ultracool dwarfs of interest, we are conducting
a proper motion search that combines the first epoch of PS1 data with
2MASS, with initial results from PS1 commissioning data in
\citet{deacon11-ps1-tdwarfs}.
In this Letter, we present the discovery of a very high proper motion
late-T~dwarf, \psobjectfull\ (hereinafter \psobject). This object has
independently been identified by \citet{2011A&A...532L...5S} in a search
for bright, very red objects in the \WISE\ Preliminary Data Release with
proper motions of $\gtrsim$0.3\arcsec/year. They estimated a spectral
type of T8--T10 based on the $(W1-W2)$ color, a photometric distance of
5.5$^{+2.3}_{-1.6}$~pc, and possible thick disk membership. Our near-IR
spectroscopy gives a spectral type of T8, a photometric distance of
$7.2\pm0.7$~pc, and probable thin disk membership.

\section{Mining \ps, 2MASS, and WISE}

The \PS\ survey is obtaining multi-epoch imaging in 5~optical bands
(\grizy) with a 1.8-meter wide-field telescope on Haleakala, Maui.
Images are processed nightly through the Image Processing Pipeline (IPP;
\citealp{PS1_IPP, PS1_photometry, PS1_astrometry}). Photometry is on the
AB magnitude scale, and the first generation of astrometry is tied to
2MASS, with limiting uncertainties of $\approx$70~mas.
Although the PS1 filter system \citep{PS1_lasercal} is very similar to
SDSS \citep{1996AJ....111.1748F}, there are differences. Most notable
are the facts that the \zps\ filter is cut off at 840~nm unlike the
long-pass $z_{SDSS}$ filter, and SDSS has no corresponding \yps\ band
($\lambda_C=990$~nm, $\Delta\lambda=70$~nm), which exploits the
excellent red sensitivity of the PS1 detectors.\footnote{See also {\tt
    http://svn.pan-starrs.ifa.hawaii.edu/trac/ipp/wiki/PS1\_Photometric\_System}.}

We use data from the \TPS, the most time-intensive of the several PS1
surveys. This survey covers the 3$\pi$ steradians north of declination
$-$30\degs, with each filter observed for a total of 6~epochs over
3~years, and with two exposures at each epoch spaced $\sim$30~min apart
to identify solar system objects.
We used the the IPP software known as DVO (Desktop Virtual Observatory;
\citealp{PS1_IPP}) for managing the large numbers of PS1 detections and
enabling large-area cross-correlations ($\sim$10$^{10}$~objects). We
ingested 2MASS into the DVO database to mine the two datasets.
At the time we did this (February 2011), PS1 had not yet completed its
first sweep of the sky. Since the filters for the \TPS\ are observed
weeks to months apart to optimize sky brightness conditions and parallax
factors, we maximized our search area by focusing on the \yps\ data
alone, rather than employing multi-color criteria covering a smaller
patchwork on the sky.

Our search method is described in detail by
\citet{deacon11-ps1-tdwarfs}. In brief, we identified proper motion
candidates by matching PS1 \yps\ detections with 2MASS $J$ band
detections, excluding galactic latitudes of $|b| < 10\degs$ and allowing
PS1-2MASS separations of $\le$28\arcsec.
The PS1 data were taken from June~2009 to January~2011, during
commissioning and the first year of operations; as 2MASS was in
operation from 1997--2001, our search has a mean time baseline of
11~years, leading to a maximum proper motion of $\approx$2.7\arcsec/yr.
We used generous PS1-2MASS color cuts to remove M~dwarfs and some
instrumental artifacts.
False associations, whereby matches between PS1 and 2MASS detections in
fact are two distinct stationary objects, were then screened using
USNO-B \citep{2003AJ....125..984M} and SuperCOSMOS
\citep{2001MNRAS.326.1279H}, to depths of $R<20.5$ and $I<19$~mag.
(A separate check of objects with \mbox{$I=16-19$~mag} counterparts
found no strong T~dwarf candidates.)
Our search reaches magnitudes of $J\approx16.6$~mag and
$\yps\approx19.6$~mag, with 2MASS setting the search depth for T~dwarfs
(\eg, Figure~10 of \citealp{2009ApJ...704.1519D}).
Given our limiting magnitudes, our culling process is inevitably
incomplete. Indeed, the number of candidates increases at larger proper
motions as expected due to the remaining mispairs.

Our core PS1+2MASS search program focuses on objects with
$\lesssim$0.8\arcsec/year, where the false alarm rate is tractable.
However, the release of the \WISE\ Preliminary Data Release
\citep{2010AJ....140.1868W} in May~2011 opened the door to mining higher
proper motions, by adding unique color information and an epoch of
astrometry from the first half of 2010. We cross-matched our 21,070 high
proper motion candidates against the \WISE\ data at the PS1 position and
excluded objects with a \WISE\ source at the 2MASS position, returning
8,140 matches. For the 131~objects with $(W1-W2)>0.7$~mag, corresponding
to spectral type T0 and later, we screened the sample visually to excise
image artifacts (using data from all the input surveys) and remaining
faint mispairs (using the DSS $R$-band images, as bright early-T~dwarfs
might have counterparts on the $I$-band plates).
For the remaining 58~objects, the separations between the PS1 and \WISE\
positions showed the expected bimodal distribution of small separations
(true matches) and large separations (false matches), with a minimum
around 4\arcsec. Thus, we required the two positions to agree within
4\arcsec, leaving 49~objects.

\begin{figure}
\hbox{\hskip 0.5in
\includegraphics[width=3.5in,angle=90]{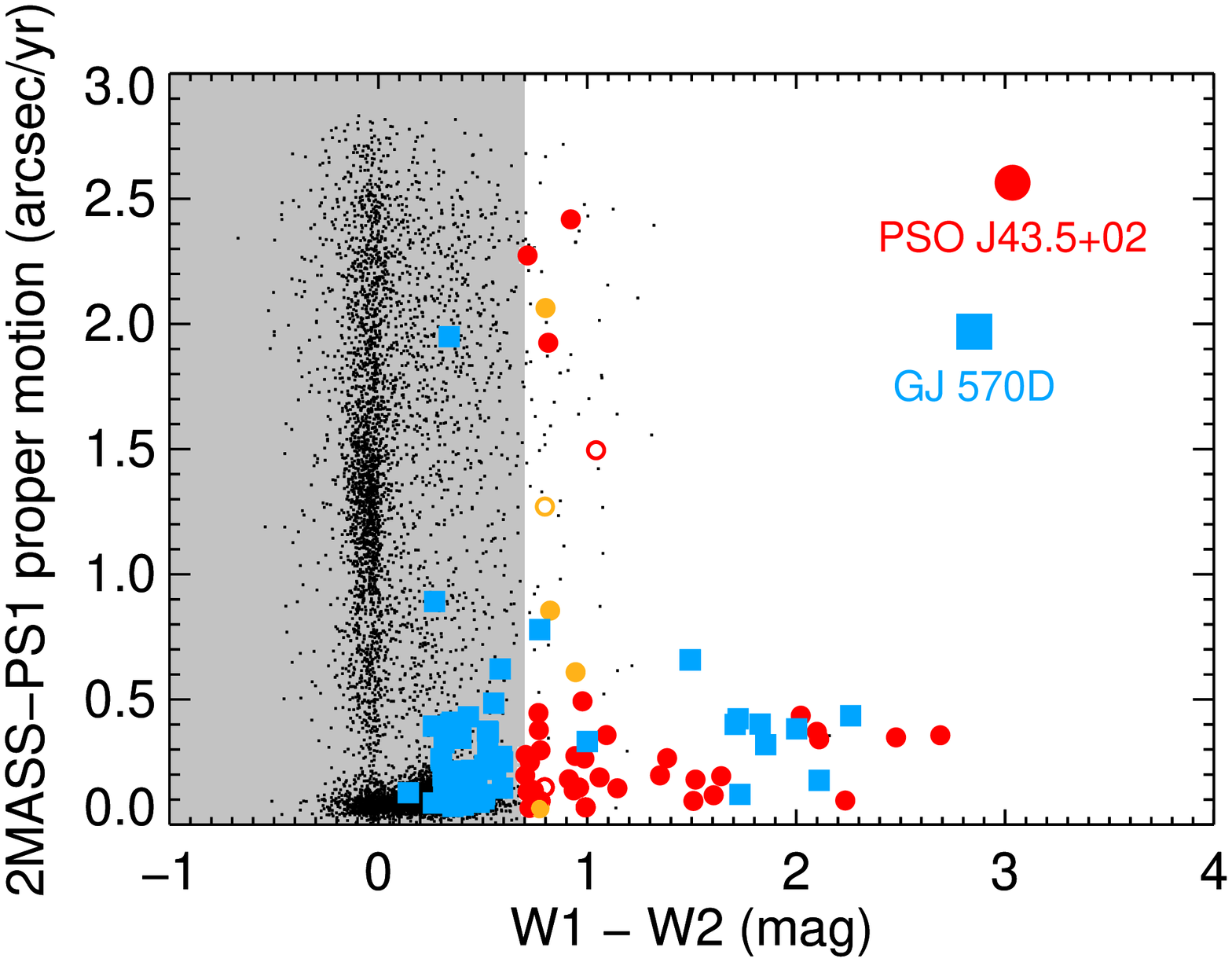}}
\hbox{\hskip 0.5in
\includegraphics[width=3.5in,angle=90]{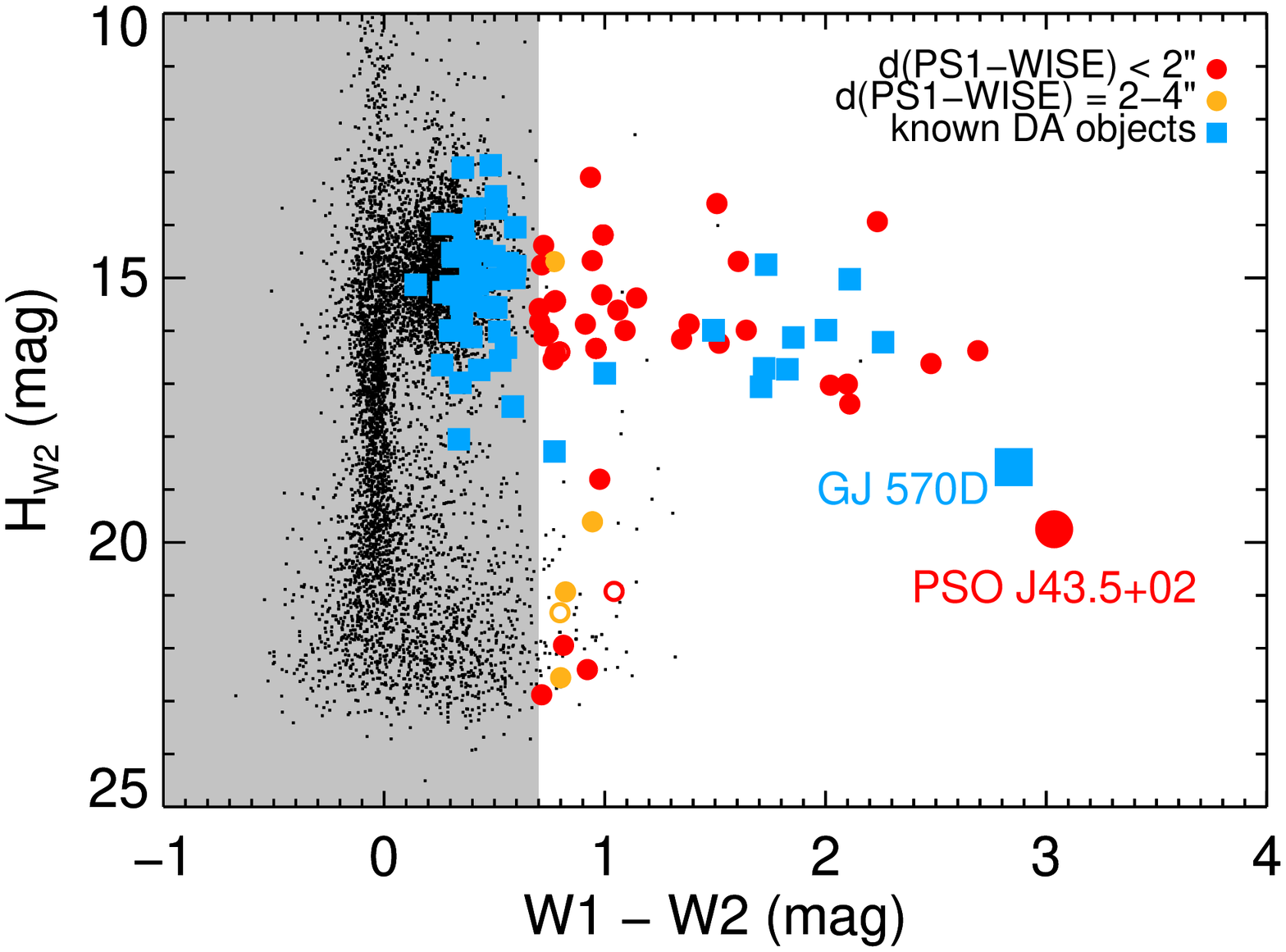}}
\caption{\normalsize Our PS1+2MASS candidates as a function of \WISE\
  color, proper motion, and $W2$ reduced proper motion. The black dots
  show our set of potential PS1+2MASS matches; at high proper motion,
  most are expected to be false associations. Objects with $(W1-W2)>0.7$
  were further screened, and the red and gold dots show those with
  \WISE\ counterparts close to the PS1 position. Objects with open
  symbols have PS1 optical counterparts at \gps\ and/or \rps. Known
  L~and T~dwarfs from Dwarfarchives.org are plotted as blue
  squares. \label{fig:selection}}
\end{figure}

The final sample is illustrated in Figure~\ref{fig:selection}. Two
objects stand out as being very red in $(W1-W2)$ and having high proper
motions. (Objects with proper motions of $<$0.5\arcsec/yr will be
discussed in a future paper.) One is the T7.5 dwarf GJ~570D
\citep{2000ApJ...531L..57B}. The other, \psobjectfull \footnote{The PS1
  name used here is based on the computed position at epoch 2010.0.}
(\wiseobjectfull), was previously unknown at the time. We identified a
$z_{SDSS}$-only counterpart in the latest SDSS data release (DR8;
\citealp{2011ApJS..193...29A}). This object's absence in earlier
releases explains why it was not found in past SDSS searches. Similarly,
its faint $J$ band magnitude excluded it from previous 2MASS searches.

At the time of our search, about 8200~sq.~degs had \yps\ data and were
in the \WISE\ Preliminary Data Release, with about 7800~sq.~degs of the
area also having \zps\ data. (This includes the 74\% fill factor for
individual PS1 exposures due to gaps between CCDs, masked pixels around
bright stars, and bad regions of the detectors.)
To check our completeness, we examined the known T~dwarfs from
Dwarfarchives. Out of the 16 objects that are present in \WISE, bright
enough to be in 2MASS, and residing within the area imaged by PS-1, we
retrieved 12 of them, with the rest being culled during the filtering
process described above, \eg, by being too close to a USNO-B source.
Thus our estimated completeness is 75\%.
The other high proper motion object found by
\citet{2011A&A...532L...5S}, the T9--T10~dwarf WISEP~J1741+2533, was not
selected as its sky position did not have \yps\ imaging when we did our
data-mining.

\section{Spectroscopy}

We obtained low-resolution ($R\approx$100) spectra of \psobject\ on
2011~June~25 and July~21~UT from NASA's Infrared Telescope Facility
located on Mauna Kea, Hawaii. The two spectra agree well, but the July
one had higher S/N so we use it here. Conditions were photometric with
average seeing. We used the near-IR spectrograph SpeX
\citep{1998SPIE.3354..468R} in prism mode with the 0.8\arcsec\ slit,
obtaining 0.8--2.5~\micron\ spectra in a single order. The total
on-source integration time was 16~min.
We observed the A0V star HD~18571
contemporaneously with \psobject\ for telluric calibration. All spectra
were reduced using version~3.4 of the SpeXtool software package
\citep{2003PASP..115..389V,2004PASP..116..362C}.

The spectrum shows the strong \htwoo\ and \meth\ absorption
characteristic of late-T~dwarfs (Figure~\ref{fig:spectrum}). We
classified \psobject\ using the five spectral indices from
\citet{2006ApJ...637.1067B}, with spectral types assigned using the
polynomial fits from \citet{burg2006-lt}. Excluding the saturated
\meth-$K$ index, the average spectral type from the indices was T8.0
with an RMS of 0.14~subclasses. We also computed the $W_J$ and
$\ammonia$-$H$ indices of \citet{2007MNRAS.381.1400W} and
\citet{2008A&A...482..961D}, respectively, and found values similar to
known T8 dwarfs \citep[e.g.][]{2010MNRAS.406.1885B}.

In addition, we visually classified \psobject\ by comparing with SpeX
prism spectra of late-T dwarf spectral standards. Following the
prescription of \citet{2006ApJ...637.1067B}, the depth of the \htwoo\
and \meth\ absorption bands were examined, normalizing the spectra of
\psobject\ and the standards at their $J$, $H$, and $K$-band peaks. The
agreement with the T8~standard 2MASS~J0415$-$0935 is excellent, with
\psobject\ showing stronger absorption than the T7.5~dwarf GJ~570D.
The T8.5 objects from \citet{2008A&A...482..961D}
and \citet{2008MNRAS.391..320B}
have deeper \htwoo\ absorption and narrower $J$~and $H$~band continuua
than \psobject.
Thus, the indices and visual typing both give a spectral type of T8.

We fit the solar metallicity BT-Settl-2010 model atmospheres
\citep{2010arXiv1011.5405A} to our spectrum. The models span
$\Teff=500-1500$~K ($\Delta\Teff=100$~K) and $\logg=4.0-5.5$ (cgs;
$\Delta\logg=0.5$). We first flux-calibrated our spectrum to the
weighted average of the 2MASS $J$~and $H$~band photometry (which is only
S/N~$\approx$~5--6 in each filter). Following
\citet{2009ApJ...706.1114B} and \citet{2007arXiv0711.0801C}, we used a
Monte Carlo approach to fit the 0.8--2.4~\micron\ data, excluding the
1.60--1.65~\micron\ region because of the known incompleteness of the
methane line list. For each Monte Carlo trial, we changed the flux
calibration and spectrum assuming normal distributions for the
photometric and spectroscopic uncertainties, respectively. We then fit
the model atmospheres to each artificial spectrum and repeated the
process 10$^3$ times.

The best-fitting model has $\Teff=800$~K and $\logg=4.0$
(Figure~\ref{fig:spectrum}), comparable to atmosphere fitting results
for other comparably late-type objects
\citep[e.g.][]{2006astro.ph.11062S, 2007ApJ...667..537L,
  2011arXiv1103.0014L}.
The broad-band photometry agrees well with the model, except for $W2$
where the model is a factor of $2.92\pm0.14$ fainter. 
The mid-IR colors of late-T dwarfs show significant scatter with
spectral type (Figure~\ref{fig:spectrum}), due to non-equilibrium
CO/\meth\ chemistry, metallicity, and surface gravity effects
\citep[e.g.][]{2010ApJ...710.1627L}, which are not fully incorporated
into current models. This empirical scatter also explains the spectral
type of T8--T10 inferred by \citet[][]{2011A&A...532L...5S} based on the
mid-IR colors, compared to the actual type of T8.

\section{Physical Properties of \psobject}

Based on an unweighted fit to 19~objects from L5--T10 with parallaxes
and \WISE\ data, we derive a 3rd-order polynomial for the $W2$ band
absolute magnitude $M(W2) = \sum_i\ a_i\ (SpT)^i$ where $a_i = \{-1.075,
1.580, -0.06780, 0.001066\}$ and $SpT$=15 for L5, =20 for T0, etc. The
rms scatter about the fit is only 0.18~mag.
This gives a photometric distance of $7.2\pm0.7$~pc.\footnote{Another
  distance estimate comes from the ``spectroscopic parallax'' method of
  \citet{2009ApJ...706.1114B}, whereby
the model atmosphere fitting produces a distance of
$(12.1\pm0.7)(R/\Rjup)$~pc where $R$ is the object's radius.
The Lyon/Cond evolutionary models \citep{2003A&A...402..701B} predict a
radius of 0.10--0.08~\Rsun\ for an effective temperature of 800~K and
ages of 1--10~Gyr, resulting in a spectroscopic distance of 10--12~pc.
\citet{2011arXiv1103.0014L} show that such distances for late-T~dwarfs
range from $\approx$1--2$\times$ the parallactic distance.}

The 2.56\arcsec/yr proper motion of \psobject\ is exceptional. Among
field T~dwarfs, the only object with a larger value is the T7.5~dwarf
2MASS~J1114$-$26 \citep[3.05\arcsec/yr;][]{2005AJ....130.2326T}, which
\citet{2007ApJ...667..537L} characterize as slightly metal-poor
($[M/H]\approx-0.3$~dex). Our photometric distance of $7.2\pm0.7$~pc
leads to a tangential velocity of $87\pm8$~\kms.
Following \citet{2009ApJ...699..168D}, we assess the galactic membership
of \psobject\ using an oversampled simulation of the stars within 30~pc
from the Besan\c{c}on model \citep{2003A&A...409..523R}.
The model provides \vtan\ distributions for thin disk, thick disk, and
halo stars. We compute the membership probabilities for \psobject\ from
the fraction of its nearest neighbors that belong to each model
population. For \vtan~=~87~\kms, we find membership probabilities in the
thin/thick disk of 0.87/0.13, with $<$0.01 probability of belonging to
the halo. Thus, even though the \vtan\ of \psobject\ is large, it is
merely on the cusp of thin/thick disk membership, because of the many
more thin disk stars (a divide of 0.973/0.026/0.002
between thin disk/thick disk/halo members in this model). Thin disk
membership is preferred, even accounting for the measurement
uncertainty; at the $\pm$1$\sigma$ limits of \vtan\ (79 and 95~\kms),
the thin/thick disk probabilities are 0.90/0.10 and 0.77/0.23,
respectively.\footnote{Out of the current census of
  $\approx$100~T~dwarfs with proper motion measurements, only three
  objects have \mbox{$\vtan > 100$~\kms:} 2MASS~J1114$-$26 (T7.5,
  $140\pm22$~\kms; \citealp{2009AJ....137....1F}), ULAS~J1319+1209 (T5,
  $192\pm40$~\kms; \citealp{2011MNRAS.414..575M}), and ULAS~J0926+0835
  (T4, $213\pm59$~\kms; \citealp{2011MNRAS.414..575M}). 
  Murray \etal\
  suggest their two ULAS objects are likely halo objects, based on their
  kinematics relative to the galactic disk velocity ellipsoid. Our
  approach finds thick disk membership seems plausible. At the
  $\pm$1$\sigma$ limits on \vtan, ULAS~J1319+1209 has probabilities of
  0.09/0.81/0.10 and 0.00/0.29/0.71 for thin disk/thick disk/halo
  membership, and ULAS~J0926+0835 has 0.07/0.82/0.11 and 0.00/0.15/0.85.
  Likewise, for 2MASS~J1114$-$26, we find probabilities of
  0.36/0.63/0.01 and 0.02/0.83/0.15, again suggesting thick disk
  membership.}

The near-IR spectrum indicates a metallicity comparable to other field
objects (Figure~\ref{fig:spectrum}).
\psobject\ is very similar to the T8~dwarf 2MASS~J0415$-$09, which is
inferred to have [Fe/H]~=~0.0--0.3 and $\logg=5.0-5.4$
\citep{2006astro.ph.11062S}.
A solar metallicity is also signalled by the $Y$~band continuum shape,
which is metallicity-sensitive \citep[][]{2006ApJ...639.1095B,
  2007ApJ...667..537L}. \psobject\ shows a more pointed shape than the
T7.5~dwarf 2MASS~J1114$-$26, which is inferred to be sub-solar
([Fe/H]~=~$-$0.3; \citealp{2007ApJ...667..537L}).
Finally, the weaker $K$~band flux of \psobject\ compared to the T8~dwarf
Ross~458C \citep{2010MNRAS.405.1140G, 2010A&A...515A..92S} further
indicates that \psobject\ has a metallicity and surface gravity (age)
typical of field objects. The youth (150--800~Myr) and metal-richness
([Fe/H]~=~+0.2--0.3) of Ross~458C reduce the effect of collision-induced
\htwoo\ opacity, leading to brighter $K$~band emission
\citep[e.g.,][]{2006liu-hd3651b}.

Our model atmosphere fitting gives a low surface gravity (\logg=4.0);
however, the result is ill-constrained, as seen by the $\chi^2$~surface
(Figure~\ref{fig:spectrum}). Following \citet{2007MNRAS.381.1400W}, a
better constraint comes from comparing \psobject\ to other late-T dwarfs
using the $W_J$ and $K/J$ indices, guided by model atmosphere
predictions and empirical benchmarks. In such a representation (\eg,
Figure~5 of \citealp{2011arXiv1103.0014L}), \psobject\ is inferred to
have a relatively high surface gravity ($\logg\approx5.0-5.5$) based on
its large $K/J$ value, consistent with an old ($\gtrsim$Gyr) age.

PS1 \TPS\ \zps\ and \yps\ imaging are scheduled to maximize the parallax
factor between epochs, with the goal of deriving parallaxes over the
entire survey. Given the proxmity of \psobject, a preliminary result is
possible even with only a 5-month PS1 baseline (3 epochs from 2010.65 to
2011.06, with 2 exposures per epoch), by adding astrometry from 2MASS
(2000.73), SDSS (2008.72), and \WISE\ (2010.08). Altogether, these data
span a 10-year baseline.
For \WISE, there are known $\sim$0$\farcs$5 systematic errors in the
Preliminary Release Source Catalog, so we constructed our own
astrometric catalog for 1~sq.~deg around \psobject\ from the Preliminary
Single Exposure (L1b) Working Database, which we found to be unaffected
by such problems. We cross-matched individual \WISE\ detections and used
the weighted average and standard error for positions and uncertainties.
For 2MASS and SDSS, we assumed relative positional uncertainties of
100~mas and 70~mas, respectively, based on our experience with matching
these catalogs to high-precision ($\lesssim10$~mas) relative astrometry
from our parallax program on the Canada-France-Hawaii Telescope
\citep{dupuy2010-PhD}.
We then combined all nine astrometric datapoints (2MASS, SDSS, \WISE,
and 6~$\times$~PS1), allowing for small shifts and a full linear
transformation.
We used a Markov Chain Monte Carlo method to determine the posterior
distributions of the proper motion and parallax. The best solution had
$\chi^2 = 9.9$ (13 degrees of freedom), validating our astrometric
errors. a parallax of $171\pm45$~mas (Figure~\ref{fig:parallax}), and a
proper motion of $\mu = 2\farcs559 \pm 0\farcs011$/yr and ${\rm P.A.} =
84\fdeg82 \pm 0\fdeg25$. This includes a (negligible) correction from
relative to absolute parallax of $1.34\pm0.15$~mas computed using the
Besan\c{c}on model as described in \citet{dupuy2010-PhD}.

\section{Discussion}

For the median \vtan\ of 30~\kms\ of ultracool dwarfs within 20~pc
\citep{2009AJ....137....1F}, our proper motion range of
0.5--2.7\arcsec/yr corresponds to distances of 12~pc and 2~pc. (Objects
with higher tangential velocities of course can be selected at even
larger distances.)
This approximate kinematic limit is well-matched to the
$J\approx16.5$~mag limit for our search: the three T7.5~dwarfs with
2MASS photometry and parallaxes have $M_J$~=~15.93, 15.65, and 16.47~mag
(HD~3651B, 2MASS~J1217$-$03, and GJ~570D, respectively) and the T8~dwarf
2MASS~J0415$-$09 has $M_J=16.90$~mag, meaning our search has a limiting
distance of $\approx$12~pc for T8~dwarfs (9~pc for the 2MASS full-sky
completeness of $J=15.8$). The limiting distance drops precipitously for
later types: the T10~dwarf UGPS~J0722$-$05 \citep{2010MNRAS.408L..56L}
at 4~pc has $J=16.49\pm0.13$~mag, right at the nominal 2MASS limit.

The modest number of candidates from our search suggests that the
immediate ($\sim$10~pc) solar neighborhood does not contain a large
reservoir of undiscovered T~dwarfs down to about T8. We found
7--10~candidates (depending on the cutoff used for the PS1-\WISE\
separation) with proper motions of 0.5--2.7\arcsec/year and colors of
$(W1-W2)>0.7$, including 3~previously known objects. In comparison,
Dwarfarchives.org lists 30 objects in the same proper motion range and
with spectral types of T0--T8 (ignoring 2 tight companions found by
adaptive optics imaging). Since our PS1+2MASS+\WISE\ search covered
$\approx$20\% of the sky, a rough estimate based on the known census
predicts we should identify about 6~objects, consistent with our
findings.

\acknowledgments

The Pan-STARRS1 surveys have been made possible by the Institute for
Astronomy, the University of Hawaii, the Pan-STARRS Project Office, the
institutions of the Pan-STARRS1 Science Consortium ({\tt
  http://www.ps1sc.org}), and NASA.
Our research has employed the \WISE, SDSS-III, and 2MASS data products;
NASA's Astrophysical Data System; the SIMBAD database;
the M, L, and T~dwarf compendium housed at DwarfArchives.org,
and the SpeX Prism Spectral Libraries.
This research was supported by NSF grants AST-0507833 and AST09-09222
(awarded to MCL), AST-0709460 (awarded to EAM), AFRL Cooperative
Agreement FA9451-06-2-0338, and DFG-Sonderforschungsbereich 881 "The
Milky Way".
Finally, the authors wish to recognize the very significant cultural
role that the summit of Mauna Kea has always had within the indigenous
Hawaiian community. We are most fortunate to conduct observations from
this mountain.

{\it Facilities:} \facility{IRTF (SpeX), PS1 (GPC1)}

\clearpage


\begin{figure}
\hskip 0.5in
\includegraphics[height=5.5in,angle=0]{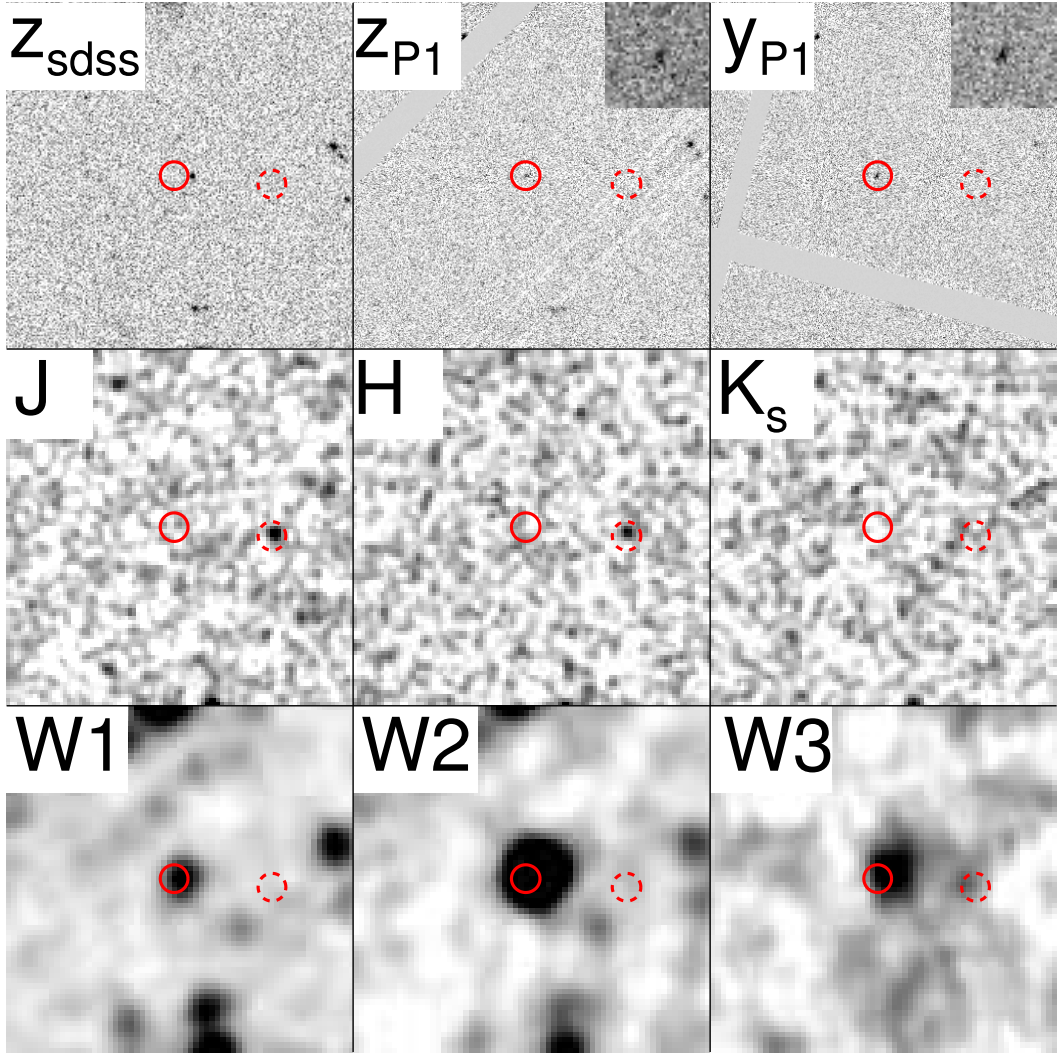}
  \caption{\normalsize Finding charts for \psobject\ from SDSS (epoch
    2008.72), \PS\ (2010.67 for \zps\ and 2010.65 for \yps), 2MASS
    (2000.73), and \WISE\ (2010.08). The red circles show the object's
    position at the 2MASS epoch (dotted line) and the PS1 \yps\ epoch
    (solid line). The PS1 inset images are 10\arcsec\ on a
    side. \label{fig:finder}}
\end{figure}

\begin{figure}
\vskip -1.2in
\hskip 0.82in
\includegraphics[width=3.72in,angle=90]{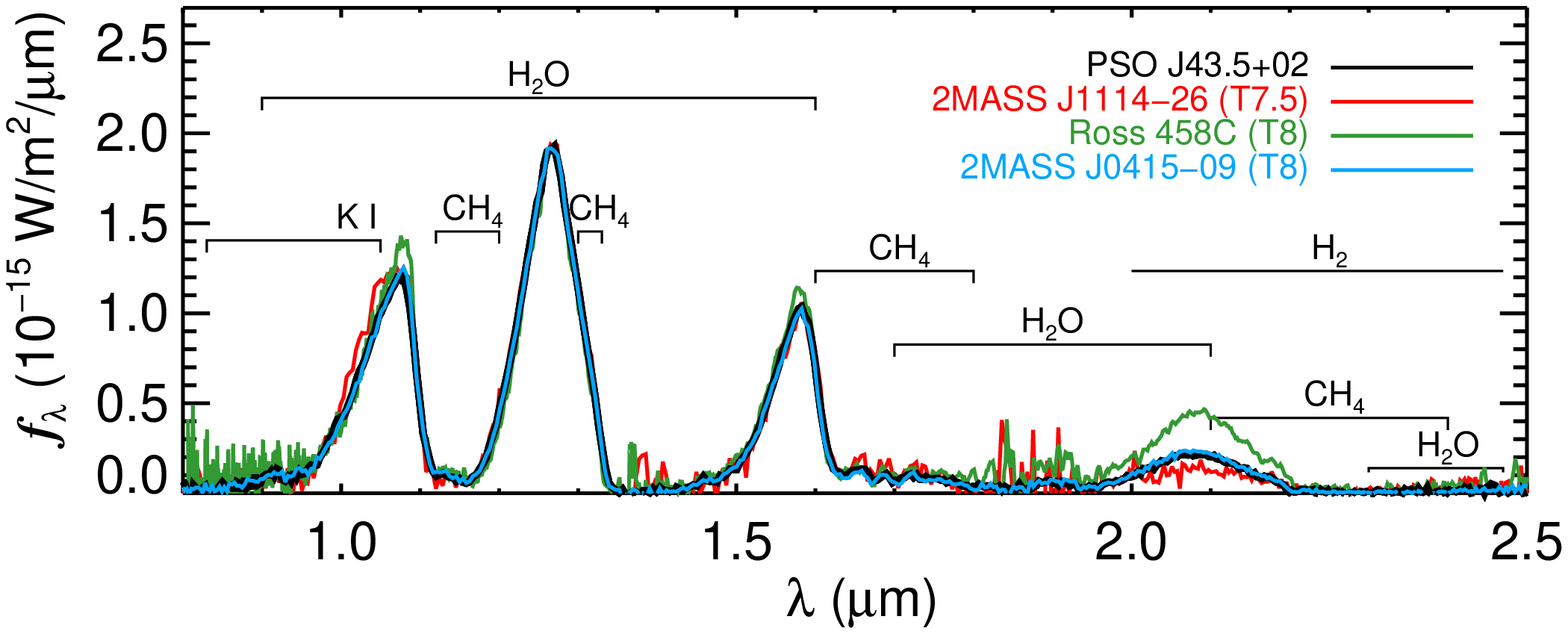}
\vskip -1.5in
\hskip 0.43in
\includegraphics[width=5.9in,angle=0]{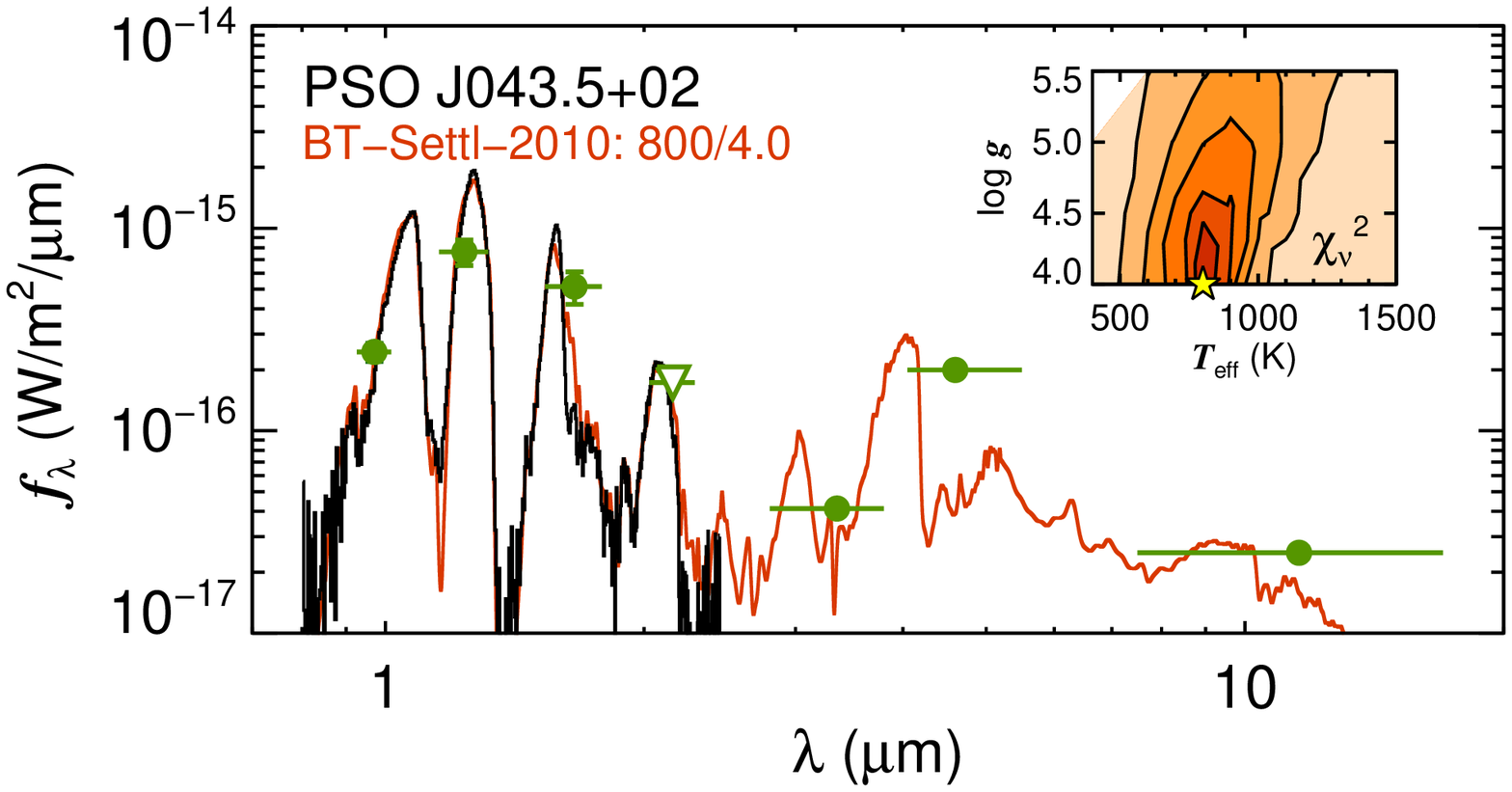}
\vskip -1in
\hskip 0.35in
\includegraphics[height=2.7in,angle=0]{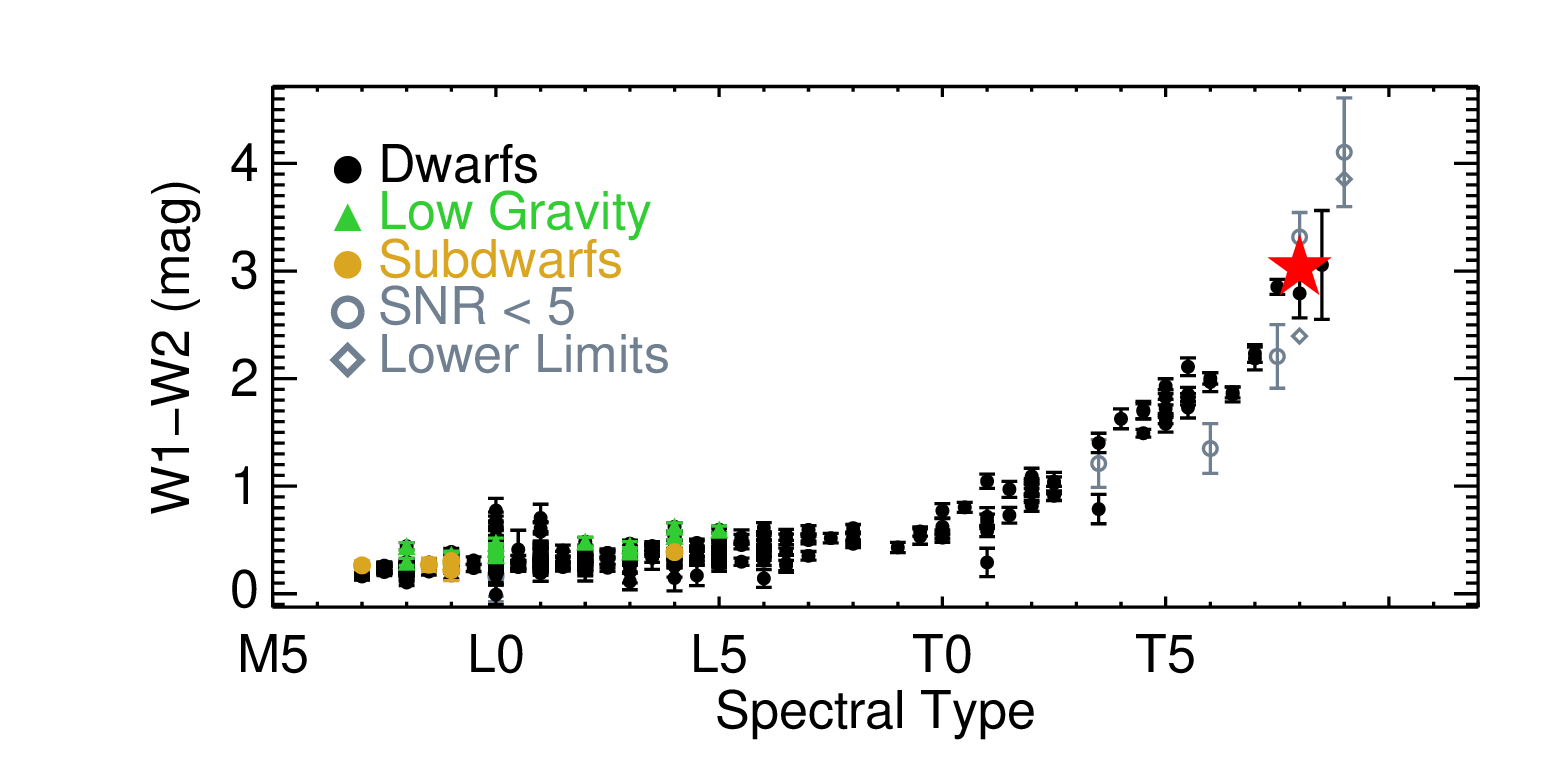}
\vskip -0.2in
\caption{\normalsize Spectral properties of \psobject. {\bf
    Top:}~Spectrum (black line) compared to other T7.5--T8 dwarfs 
  \citep{2004AJ....127.2856B, 2006ApJ...639.1095B, 2010ApJ...725.1405B},
  normalized to the 1.26--1.27~\micron\ region. The Ross~458C spectrum
  has been lightly
  smoothed.
  {\bf Middle:}~Spectrum compared to the PS1, 2MASS, and \WISE\
  photometry (green dots; Table~\ref{table}) and the best-fitting
  BT-Settl-2010 model atmosphere ($\Teff=800$~K, $\logg=4.0$~dex; orange
  line). The inset contour plot shows the mean $\chi^2$ surface from our
  model fitting. The unfilled green triangle shows the 2MASS \Ks~band
  upper limit.
  {\bf Bottom:}~\WISE\ $(W1-W2)$ color as a function of spectral type,
  with subsets from \citet{2009AJ....137....1F} highlighted
  in color. This plot shows objects from Dwarfarchives.org with good
  (unconfused and not blended) \WISE\ data and additional objects from
  \citet{2011ApJ...726...30M} and \citet{2011arXiv1106.3142G}. Known
  binaries and close companions are excluded. When available, optical
  spectral types are used for M~and L~dwarfs, with near-IR types for
  T~dwarfs. \psobject\ is the red star. 
  \label{fig:spectrum}}
\end{figure}

\begin{figure}
\vskip -1in
\includegraphics[width=3in,angle=0]{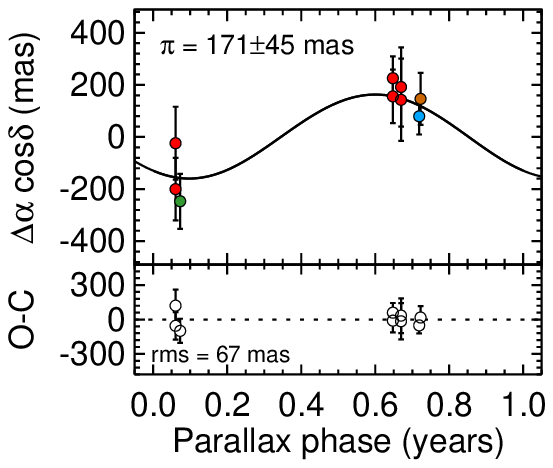}
\hskip 0.4in
\includegraphics[width=3in,angle=0]{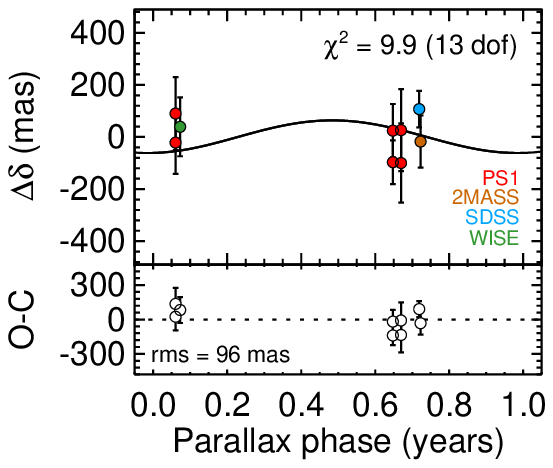}
\caption{\normalsize Preliminary parallax for \psobject. Top panels show
  relative astrometry in~$\alpha$ and~$\delta$ as a function of Julian
  year after subtracting the best-fit proper motion. (This
  representation is for display purposes only. Our analysis fits for
  both the proper motion and parallax simultaneously.) Bottom panels
  show the residuals after subtracting the parallax and proper motion.
  \label{fig:parallax}}
\end{figure}

\clearpage

\begin{deluxetable}{lccccccc}
\tablecaption{Measurements of \psobject\ \label{table}}
\tabletypesize{\small}
\tablewidth{0pt}
\tablehead{
  \colhead{Property} &
  \colhead{Measurement} 
}

\startdata
\cutinhead{Astrometry\tablenotemark{a} (Equinox 2000)}
2MASS RA, Dec (ep 2000.727)          & 02:54:07.886, +02:23:56.35 \\ 
SDSS       RA, Dec  (ep 2008.723)  &  02:54:09.242, +02:23:58.32 \\
PS1 \yps\  RA, Dec  (ep 2010.651)  &  02:54:09.575, +02:23:58.68 \\
PS1 \yps\  RA, Dec  (ep 2010.651)  &  02:54:09.579, +02:23:58.56 \\
PS1 \zps\  RA, Dec  (ep 2010.673)  &  02:54:09.578, +02:23:58.69 \\
PS1 \zps\  RA, Dec  (ep 2010.673)  &  02:54:09.581, +02:23:58.56 \\
PS1 \yps\  RA, Dec  (ep 2011.064)  &  02:54:09.621, +02:23:58.72 \\
PS1 \yps\  RA, Dec  (ep 2011.064)  &  02:54:09.633, +02:23:58.83 \\
WISE       RA, Dec  (ep 2010.075)  &  02:54:09.449, +02:23:58.56 \\
Proper motion amplitude $\mu$ (\arcsec/yr)  &  $2.559\pm0.011$    \\  
Proper motion PA (\degs)                    &  $84.82\pm0.25$     \\ 
Parallax $\pi$ (mas)                        &  $171 \pm 45$       \\
\cutinhead{Photometry} 
SDSS $z$ (AB mag)       & 19.86 $\pm$ 0.07    \\
PS1 \zps\ (AB mag)      & 21.25 $\pm$ 0.06\tablenotemark{b} \\
PS1 \yps\ (AB mag)      & 19.11 $\pm$ 0.04\tablenotemark{b} \\  
2MASS $J$ (mag)         & 16.56 $\pm$ 0.16    \\
2MASS $H$ (mag)         & 15.88 $\pm$ 0.20    \\
2MASS $\Ks$ (mag)       & $>$16.0 (2$\sigma$) \\
WISE $W1$ (mag)         & 15.74 $\pm$ 0.07    \\
WISE $W2$ (mag)         & 12.71 $\pm$ 0.03    \\
WISE $W3$ (mag)         & 11.04 $\pm$ 0.13    \\
WISE $W4$ (mag)         & $>$9.1  (2$\sigma$) \\
\cutinhead{Spectrophotometry\tablenotemark{c}}
MKO (synth) $J$ (mag)       &     16.14 $\pm$ 0.12 \\
MKO (synth) $H$ (mag)       &     16.51 $\pm$ 0.12 \\
MKO (synth) $K$ (mag)       &     16.84 $\pm$ 0.12  \\
MKO (synth) $(J-H)$ (mag)   &  $-$0.368 $\pm$ 0.002 \\
MKO (synth) $(H-K)$ (mag)   &  $-$0.336 $\pm$ 0.005 \\
MKO (synth) $(J-K)$ (mag)   &  $-$0.704 $\pm$ 0.005 \\
2MASS (synth) $J$ (mag)        & 16.43 $\pm$ 0.12     \\
2MASS (synth) $H$ (mag)        & 16.47 $\pm$ 0.12     \\
2MASS (synth) $\Ks$ (mag)      & 16.69 $\pm$ 0.12     \\
2MASS (synth) $(J-H)$ (mag)    & $-$0.044 $\pm$ 0.002 \\
2MASS (synth) $(H-\Ks)$ (mag)  & $-$0.213 $\pm$ 0.005 \\
2MASS (synth) $(J-\Ks)$ (mag)  & $-$0.257 $\pm$ 0.004 \\
\htwoo-$J$   &  0.039 (T8.2)  \\   
\meth-$J$    &  0.192 (T7.9)  \\   
\htwoo-$H$   &  0.175 (T8.0)  \\   
\meth-$H$    &  0.121 (T7.9)  \\   
\meth-$K$    &  0.059 ($>$T6) \\   
$W_J$        &  0.321 (T8)    \\
\ammonia-$H$ &  0.625         \\
$K/J$        &  0.121         \\
Spectral type                  & T8 \\
Photometric distance (pc)      & $7.2\pm0.7$ \\
\enddata

\tablenotetext{a}{The SDSS, PS1, and \WISE\ astrometry has been tied to
  common system, using 2MASS as the absolute reference (Section~4).}

\tablenotetext{b}{Average of multi-epoch photometry.}

\tablenotetext{c}{Broadband photometry was synthesized from our
  flux-calibrated spectrum, with errors derived from the 2MASS $J$~and
  $H$~band photometry and the spectrum's uncertainties. The errors in
  the synthesized colors are smaller than for the magnitudes, because
  the colors incorporate only the spectrum's uncertainties. Note that
  the synthesized $(J-H)$ 2MASS color ($-0.044\pm0.002$~mag) is
  consistent with the T8 spectral type, but different than the
  $0.7\pm0.3$~mag from the 2MASS catalog, likely due to underestimated
  errors in the 2MASS photometry.}

\end{deluxetable}

\end{document}